# Correlation between inherent structures and phase separation mechanism in binary mixtures

## Sarmistha Sarkar and Biman Bagchi*

Solid State and Structural Chemistry Unit, Indian Institute of Science, Bangalore – 560012, India.
Email: bbagchi@sscu.iisc.ernet.in

## *Abstract*

Binary mixtures are known to phase separate via both nucleation and spinodal decomposition, depending on the initial composition and extent of the quench. Here we develop an energy landscape view of phase separation and non-ideality in binary mixtures by exploring their potential energy landscape (PEL) as functions of temperature and composition, by employing a molecular model that promotes structure breaking abilities of the solute-solvent parent binary liquid at low temperatures. PEL that provides the inherent structure (IS) of a system is obtained by removing the kinetic energy (including that of intermolecular vibrations). Broad Distribution of inherent structure energy demonstrates the larger role of entropy in stabilizing the parent liquid of the structure breaking type of binary mixtures. *At high temperature, although the parent structure is homogenous, the corresponding inherent structure is found to be always phase separated, with a density pattern that exhibits marked correlation with the energy of inherent structure.* Over broad range of intermediate inherent structure energy, *bicontinuous phase separation prevails with interpenetrating stripes* as signatures of spinodal decomposition. At low inherent structure energy, the structure is largely phase separated with one interface where as at high inherent structure energy, we find nucleation type growth. Interestingly, at low temperature, the average inherent structure energy ($<E_{IS}>$) exhibits a drop with temperature which signals onset of crystallization in one of the phases while the other remains in the liquid state. The non-ideal composition dependence of viscosity is anti-correlated with average inherent structure energy.

# I. Introduction

The origin of the widespread and diverse non-ideal behavior of binary mixtures is commonly explained in terms of structure making or promoting and structure breaking abilities of the solute-solvent binary system [1,2]. All earlier simulations on binary mixtures were carried out mainly with *structure promoting* constituents. Such a mixture can be easily made into a good glass former by tailoring the potential [3-6]. Characteristics of the energy landscape of structure breaking liquids could be quite different from those binary mixtures which are formed by structure forming liquids [7,8].

Inherent structure analysis to explore energy landscape view of structure breaking binary mixture *has not been carried out at all*. Inherent structures are the structures of the parent liquid at local potential energy minima and are identified in simulation by a steepest descent minimization of the potential energy. These are obtained by removing the kinetic energy, including vibrations, of atoms and molecules comprising the system. Since the original pioneering work of Stillinger and Weber [9], inherent structure analysis has played an immensely important role in understanding structure and slow dynamics of supercooled liquids and glasses [10], liquid crystals [11], to name a few in the field of soft condensed matter science [9-14]. This approach to area of disordered systems has come to be known as the energy landscape view or paradigm. Many sophisticated theoretical studies are based on this energy landscape view [13, 14]. In this article, we present an energy landscape analysis of phase separation and non-ideality in structure breaking binary mixtures by calculating the energy distribution of inherent structures as a function of composition and temperature.

The kinetics of phase separation in structure breaking liquids is known to proceed through two distinct mechanisms [15-20]. In the metastable region of the phase diagram, the phase separation occurs through nucleation while in the unstable region it proceeds through spinodal decomposition [18-20]. The origin and mechanism of the phase separation in binary mixtures has been a subject of great interest in recent times [21-24]. The composition fluctuations that lead to spinodal decomposition are long ranged and different from the local fluctuations that give rise to nucleation. Kinetics of phase separation by spinodal decomposition is often described by Cahn-Hillard-Langer-Baron-Miller (CHLBM) theories [16, 17]. These theories consider the stability of the homogeneous system to infinitesimal composition fluctuation, quantified by the position and time dependent order parameter $C(r,t)$ defined as,

$$C(r,t) = x_A(r,t) - x_B(r,t),$$
(1)

where $x_A(r,t)$ and $x_B(r,t)$ are the mole fractions for solvent and solute, respectively. Phase separation subsequent to quench is driven both by the free energy surface at the quenched temperature and dynamics of diffusion processes.

The free energy surface employed in CHLBM theories is coarse grained and therefore cannot do full justice to the effects of local heterogeneity or molecular level arrangements. Some local fluctuations and molecular arrangements could be energetically more stable than the others and can have lower entropy and therefore can evolve at a slower rate than some arrangements which are in the opposite limit of energy-entropy combination. Therefore, even though equally favorable by free energy considerations, the subsequent time evolution to the phase separated state can occur at a faster speed than the former. Recent studies have explored the role of such fluctuations in condensation, crystallization and nano particle growth [25-27]. These compositional fluctuations are therefore particularly important in the initial kinetics of phase separations.

A related problem, also dependent on spontaneous composition fluctuations, is the origin of pronounced non-ideality exhibited by many binary mixtures. Despite the great importance of non-ideality and phase separation in binary mixtures, we are yet to evolve a molecular level understanding of both the phenomena. Neither the well-known lattice model approach nor the recent mode coupling theory analysis [2] provides any satisfactory correlation between composition fluctuation and phase separation kinetics.

In this work, we present detailed analysis of inherent structure of binary mixtures to study correlation between the inherent structure energy distribution and nature of phase separation kinetics. While the parent structure is homogeneous at high temperatures, its inherent structure is always phase separated. We find that the inherent structures of structure breakers exhibit a wide range of structures ranging from nucleation to spinodal decomposition. Most importantly, *we find that there is a strong correlation between the energy of the inherent structure and the nature of incipient phase separation in the mixture.*

The organization of the rest of the paper is as follows. In section II, we describe simulation details for present model binary system. In section III, we present the correlation between non-ideality and average inherent structure energy. Section IV describes structural patterns at different inherent structure energy domains. In section V, we present quantification of spinodal decomposition and in Section VI, we describe microscopic characterization of parent liquid and the corresponding inherent structure. Section VII presents correlation between instantaneous molecular arrangements in parent liquid and the corresponding inherent structure at different temperatures. Section VIII concludes with a discussion on the significance of the results.

## II.    Simulation details

### A.    Details of the model binary systems

Our model binary system consists of total 500 particles [solvent (A) + solute (B)] enclosed in a cubic box and periodic boundary conditions were applied. In the present study we emphasize on structure breaker type of binary liquid. All three interactions such as solute-solute, solvent-solvent and solute-solvent are described by the Lennard-Jones (12-6) potential,

$$U_{ij} = 4\varepsilon_{ij}[(\frac{\sigma_{ij}}{r_{ij}})^{12} - (\frac{\sigma_{ij}}{r_{ij}})^6].\tag{2}$$

Here $i$ and $j$ denote any two particles. For simplicity, diameter ($\sigma$) and mass (m) for both solute and solvent atoms have been set to unity. The interaction strengths are $\varepsilon_{AA} = 1.0$, $\varepsilon_{BB} = 0.5$ and $\varepsilon_{AB} = 0.3$. In our present binary model, the Lennard-Jones interaction strength between A and B i.e. $\varepsilon_{AB} = 0.3$ which is less than either $\varepsilon_{AA}$ or $\varepsilon_{BB}$ and it leads to phase separation at low temperature and hence it corresponds to "structure breaking" (SB) binary mixture. For "structure promoting" (SP) type $\varepsilon_{AB}$ (> 1.5, typically 2.0) is greater than either $\varepsilon_{AA}$ or $\varepsilon_{BB}$.

TABLE I. Interaction strengths between solent(A) and solute(B) for structure promoting & structure breaking binary mixtures are shown below.

==============================================================

|                    | Structure promoter | Structure breaker |
| ------------------ | ------------------ | ----------------- |
| $\varepsilon_{AA}$ | 1.0                | 1.0               |
| $\varepsilon_{AB}$ | 2.0                | 0.3               |
| $\varepsilon_{BB}$ | 0.5                | 0.5               |

==============================================================

Advantage of our present model is that it can serve as starting point to understand many mysterious properties of binary mixtures. Reduced temperature $T^* = k_B T / \varepsilon_{AA}$ has been set higher i.e. $T^* = 1.6$ for structure breaking model than the temperature ($T^* = 1.0$) of structure promoting model. The reason behind consideration of high temperature for structure breaker is that the system phase separates (in parent structure at low temperature) and we want the parent structure to be homogeneous. We have selected an integration time step $\Delta t = 0.001\tau$ where the reduced time $\tau$ is defined as $\tau = \sigma \sqrt{\dfrac{m}{\epsilon_{AA}}}$. We have dealt with different solute compositions from 0.0 to 0.1 in steps of 0.1. For each solute composition the system has been equilibrated for 500000 steps. We carry out simulations for another 2 million steps after the equilibration and calculate all the relevant properties. The primary motivation has been to understand the origin of pronounced non-ideality in our present binary mixture and this non-ideality has always been studied as a function of A and B compositions. That is why we study the system as a function of the A and B composition. This is standard in the binary mixture literature.

Recent theories and simulations have revealed that surface tension is particularly sensitive to the range of intermolecular interactions [28]. In 2D L-J systems the surface tension of planar interface is only 0.05 (in reduced units) with a cutoff of 2.5 $\sigma$, but increases to 0.18 without the cutoff. Similarly, in a 3D system, the value 0.49 of surface tension with 2.5 $\sigma$ cut-off changes to 0.94 when the full range is included in calculations. Therefore such a strong dependence on cut-off indicates that one needs to be extra careful and should not retain just the nearest neighbor interactions. In our structure breaker model we have considered the interactions among all particles. However, we have also studied inherent structure analysis for a comparatively bigger system with 2048 particles, employing nearest neighbor list. The results show no significant differences in phase separation patterns (see the supplementary material section **S4**).

## B. Determination of viscosity

We obtain shear viscosity from integration of the stress-stress autocorrelation function using the following expression,

$$\eta = (VK_BT)^{-1} \int_0^\infty dt < \sigma^{xz}(t)\sigma^{xz}(0) >,$$  (3)

where $\sigma^{xz}$ is the off-diagonal element of the stress tensor given by $\sigma^{xz} = m\sum_{i=1}^{N}\dot{x}_i\dot{z}_i + \frac{1}{2}\sum_{i=1}^{N}F_{ij}^{Z}x_{ij}$, where $F_{ij}^{z}$ is the z component of force exerted by particle "$j$" on "$i$" and $x_{ij}$ is the x component of separation of the two particles.

## C. Computation of inherent structures

The configurations of the system corresponding to the local potential energy minima, known as the inherent structures are obtained from computer simulations. As usual, we have determined inherent structures by quenching the equilibrium configurations to their local minima using conjugate gradient (CG) algorithm [29]. CG minimizations were carried out for 2000 equilibrium configurations from which average inherent structure energies have been computed.

# III. Correlation between non-ideality and average inherent structure energy

## A. Phase diagram of the model system

Starting with high temperature, homogeneous state of the parent liquid for a particular solute composition we have carried out molecular dynamics simulations to calculate inherent structures. Gradually, the temperature of parent liquid has been decreased and several snapshots have been taken to achieve temperature of phase separation. Same procedure has been carried out for different solute mole fractions. The simulated phase diagram of the model binary mixture is shown in **Figure 1(a)**. The plot shows the familiar inverted parabolic shape in the temperature-composition plane. The figure displays a small asymmetry which is a reflection of the asymmetric interaction potential between the different species. The LJ interaction energy between A and B is assumed to be vastly different from the symmetric Berthelot's rule (which would give a value of 0.75 for $\varepsilon_{AB}$). The rule of geometric mean gives a value close to 0.7. We have done Monte Carlo simulations also. The phase diagram achieved from Monte Carlo simulation is in agreement with the present phase diagram.

## B. Temperature dependence of inherent structure energy

Molecular dynamics simulations and inherent structure analyses have been carried out on binary structure breaker liquid model with 500 particles for the first time at different mole fractions and temperatures. **Figure 1(b)** depicts the temperature dependence of the average inherent structure energy $<E_{IS}>$. The plot shows that for high temperature parent liquids, the $<E_{IS}>$ remains independent for a particular composition, suggesting that population of different inherent structures is not significantly affected. For low temperature parent liquid, however, there is a dip in the $<E_{IS}>$. This is found to be due to crystallization among the phase separated species A because near the phase boundary as in **Figure 1(a)**, both the density and temperature of the Lennard-Jones spheres lie in the solid side of the liquid-solid phase diagram. Now the effective reduced temperature of B-species is twice that of A-species ($\varepsilon_{BB} = 0.5$). Thus, the space occupied by B-species remains in the molten amorphous phase. Therefore, although the nature of the curve is similar to the ones observed in super cooled liquids [10] and liquid crystals [11], the origin of the drop of the $<E_{IS}>$ with temperature is different.

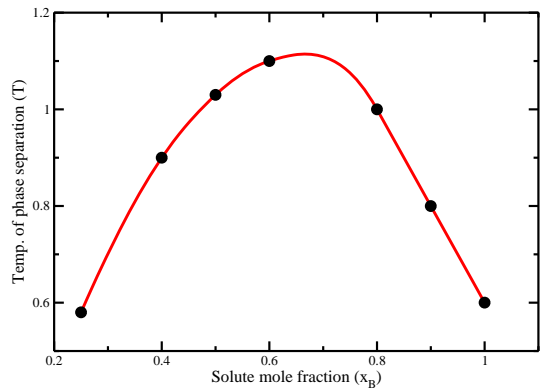

**Figure 1(a)**

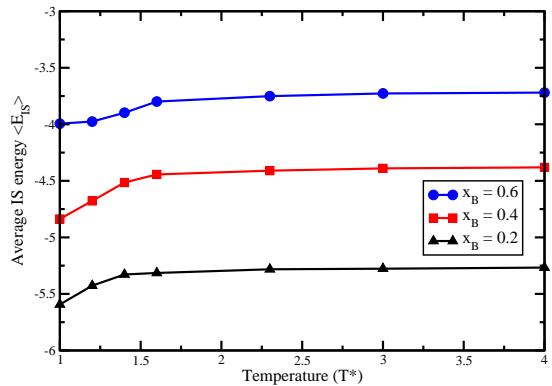

**Figure 1(b)**

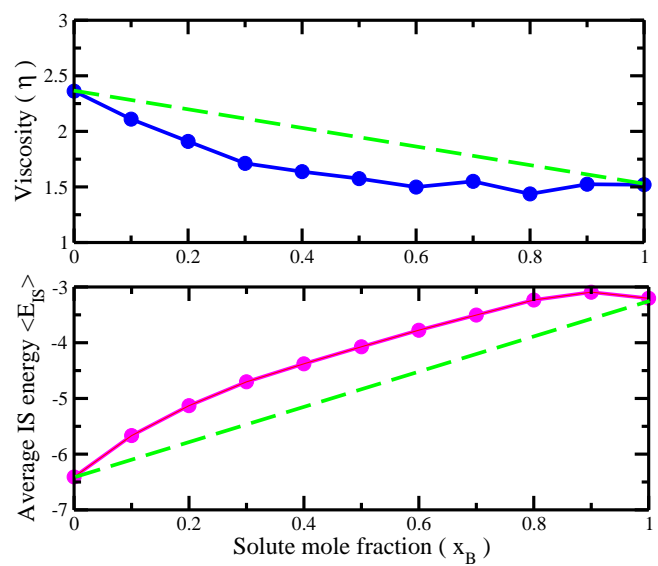

**Figure 1(c)**

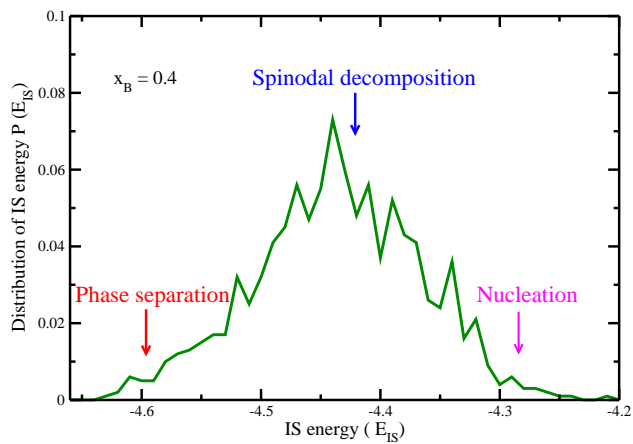

**Figure 1(d)**

**Figure 1.** **(a)** The computed phase diagram of the system. **(b)** Temperature (T*) dependence of the average inherent structure energy ( $<E_{IS}>$ ) is shown at three different mole fractions ( $x_B$ = 0.2, 0.4 and 0.6 ). **(c)** Plot of the computed viscosity and the average inherent structure energy at different solute mole fractions. Note that green lines represent the ideal Raoult's law. **(d)** The distribution of inherent structure energy sampled over 2000

parent configurations with 500 particles. The temperature of the corresponding parent liquid is T* = 1.6 and mole fraction ($x_B$) = 0.4. Note the broad distribution.

## C. Nonlinear composition dependence of inherent structure energy and its correlation with viscosity

The nonlinear composition dependence of the $<E_{IS}>$ and its correlation with viscosity are depicted in **Figure 1(c)**. The $<E_{IS}>$ shows a positive deviation from the ideal Raoult's law and the energy lies entirely above the linear line that gives the ideal behaviour. The viscosity, on the other hand, shows a clear negative deviation from ideality. We observe such anti-correlation between viscosity and $<E_{IS}>$ also in the case of structure forming liquids, where, however, the variation trend of each is exactly the opposite of the one found here for the case of structure breakers (see the supplementary material section **S1**)

## D. Distribution of inherent structure energy

**Figure 1(d)** depicts the distribution of inherent structure energy **$P$ ($E_{IS}$)** for the parent liquid with temperature (T*) = 1.6 and mole fraction ($x_B$) = 0.4. The broad inherent structure energy distribution clearly indicates a large entropic contribution towards the stability and homogeneity of the parent liquid. The standard deviation of the distribution for structure breaking liquid is found to be 0.07 in the reduced unit which is more than the width (0.05) we observe for a structure forming liquid (see supplementary material section **S2**). The inherent structure energy distribution remained rugged even after averaging over a large number (2000) of configurations.

## IV. Structural patterns at different inherent structure energy domains

The wide inherent structure energy distribution $P(E_{IS})$ needs further exploration. We studied the structural aspects at three different energy regions as shown in **Figure 1(d)** of the distribution. Snapshots of these structures are shown in **Figures 2 (a), (b), (c)** and **(d)**. These snapshots reveal interesting structural differences. The spatial density distribution is found to be strongly correlated with $P(E_{IS})$. At low inherent structure energy ( $E_{IS}$ ), beautiful string-like rings [30] form along with phase separation as shown in **Figure 2(a)**. At high $E_{IS}$, we find nucleation type growth as in **Figure 2(b)**. At intermediate values of $E_{IS}$, where $P(E_{IS})$ is peaked, the density distribution of inherent structure shows remarkable pattern of spinodal decomposition with interpenetrating stripes as in **Figures 2 (c)** and **(d)**.

The correlation between the structural patterns revealed in the inherent structure can be rationalized as follows. Area of A-B contact is least for **Figure 2(a)** which causes the surface tension contribution to the free energy to be the least and thereby inherent structure energy is minimum here compared to the situations in **Figures 2 (b), (c)** and **(d)**. The fact that inherent structure with lowest energy is largely phase separated is in agreement with the lowest free energy of the system. Since the compositions in the parent liquid are in dynamical equilibrium among themselves, low energy means low entropy and that leads the structure to become more phase separated.

The partly nucleated state obtained at high energy side of the energy distribution as in **Figure 2(b)** is also consistent because the partly nucleating state has the largest surface contact between A and B. High energy corresponds to the high entropy that favours the structures in the dispersed state. This pattern also exhibits large dispersion of the solute atoms.

At the intermediate $E_{IS}$ we find alternate density stripes showing the signature of spinodal decomposition with bicontinuous phase separating structure [31] as in **Figure 2 (c)** and **(d).**

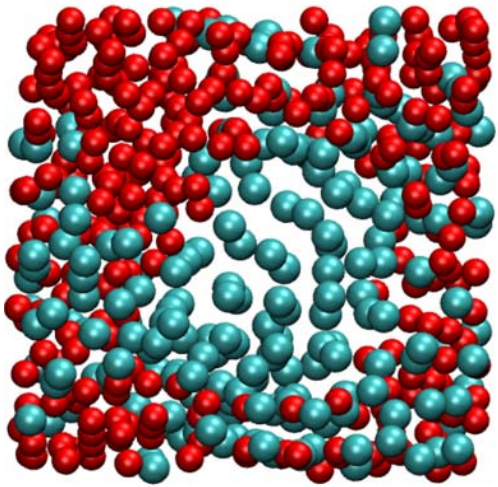

**Figure 2(a)**

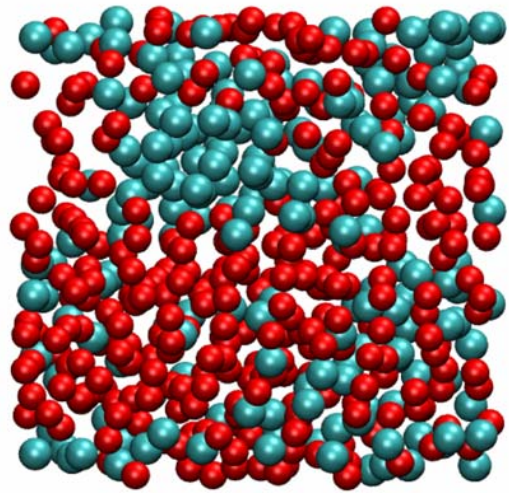

**Figure 2(b)**

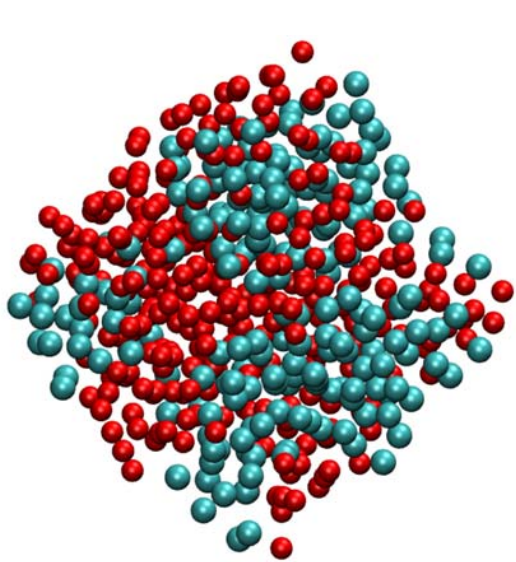

**Figure 2(c)**

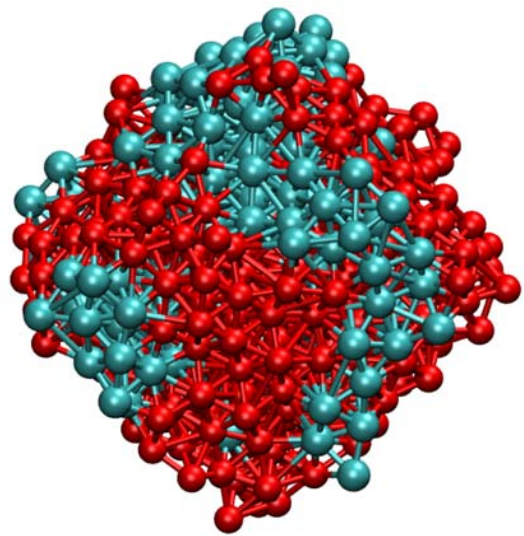

**Figure 2(d)**

**Figure 2.** Snapshots showing the spatial positions of atoms A and B in the inherent structure. The corresponding parent structure is at temperature T* = 1.6 and mole fraction ($x_B$) = 0.4. **(a)** Inherent structure corresponds to the low energy spectrum of distribution. Note the formation of beautiful string-like rings along with phase separation **(b)** inherent structure on the high energy spectrum of distribution. Note the formation of partial nucleation type structure. **(c)** and **(d)** inherent structure in the medium energy spectrum of distribution. Note the signature of spinodal decomposition as in **Figures (c) and (d)** where both snapshots are same, but in VDW & CPK mode respectively.

# V.    Quantification of spinodal decomposition

In order to quantify the extent of spinodal decomposition, we examine the density variation across several lines in different planes. The relevant order parameter is the density difference between the two species at a position r, C(r), and has already been defined by Eq. (1).

Here we show an example of variation of the order parameter across a line in **Figure 3.** Note the oscillatory variation of C(r) with respect to position (r*) reflects in a sharp peak in the structure factor S(k) that is the accepted signature of spinodal decomposition. While the oscillatory nature of variations of mole fractions $x_A$ and $x_B$ as in Figure 3 (a) and also that of C(r) with respect to r* is clear from **Figure 3 (b)**, a calculation of S(k) is not possible due to the finite size of the system.

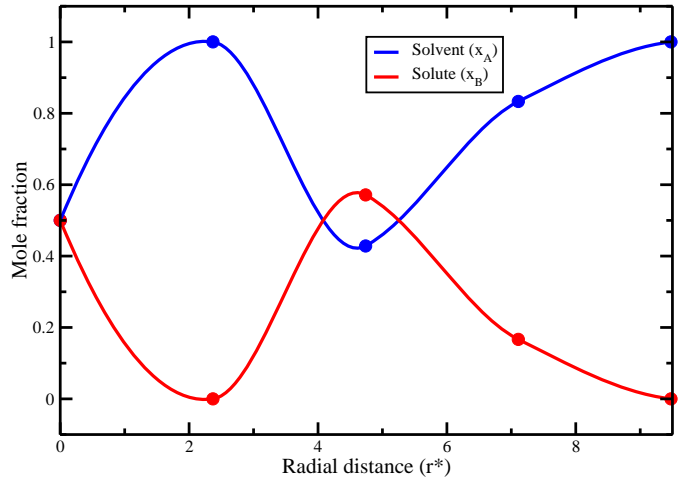

**Figure 3 (a)**

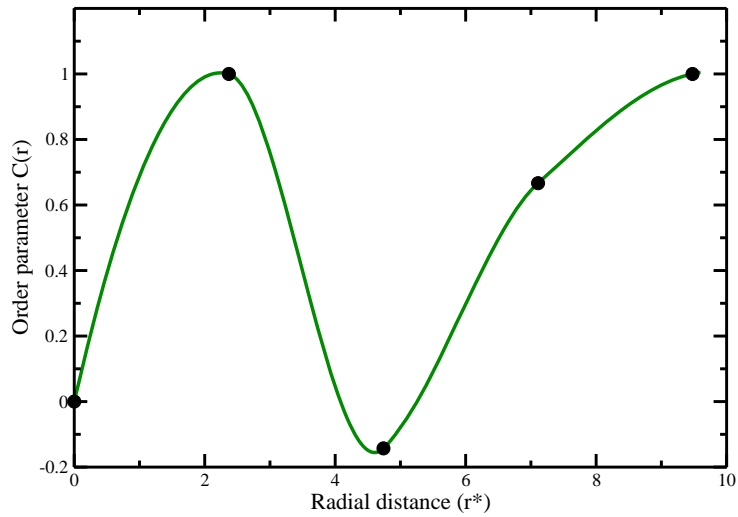

**Figure 3 (b)**

**Figure 3.** Plot of variations of mole fractions of **(a)** solvent $\left[x_A(r)\right]$ and solute $\left[x_B(r)\right]$ with radial distance (r\*); along the diagonal of a planar cross-section of the three dimensional figure. **(b)** The variation of the order parameter C(r) with radial distance (r\*). Note the oscillatory nature of the plots those represent signature of spinodal decomposition.

## VI.    Microscopic characterization of parent liquid and the corresponding inherent structure

Partial radial distribution functions for the parent liquid and corresponding inherent structure are shown in **Figures 4 (a) & (b),** respectively. These plots show that A-B correlation is the weakest compared to that of A-A and B-B. Hence, the probability of finding a particle surrounded by particles of opposite kind is less than the probability of being surrounded by particles of the same kind. In parent liquid, particularly at elevated temperatures, entropic contribution keeps the liquid homogeneous, as shown in **Figure 4(a).** In the inherent structures, the A-A correlation and B-B correlation gets substantially enhanced while that between A and B particles gets reduced. This is of course a reflection of phase separation in the inherent structure, as discussed several times in the earlier paragraphs.

**Figure 4 (b)** shows that not only there is a large sharpening of the first peak of the radial distribution functions $g^{AA}$ (r) and $g^{BB}$ (r), these correlation functions are also characterized by a split double peak which is the characteristic of amorphous packing. Similar split double peak is also seen in $g^{AB}$ (r), although the height is less in this case.

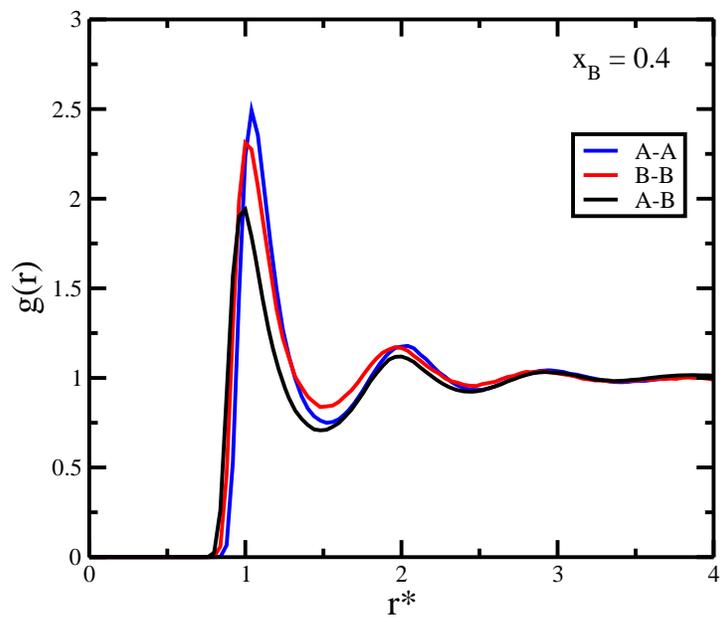

**Figure 4(a)**

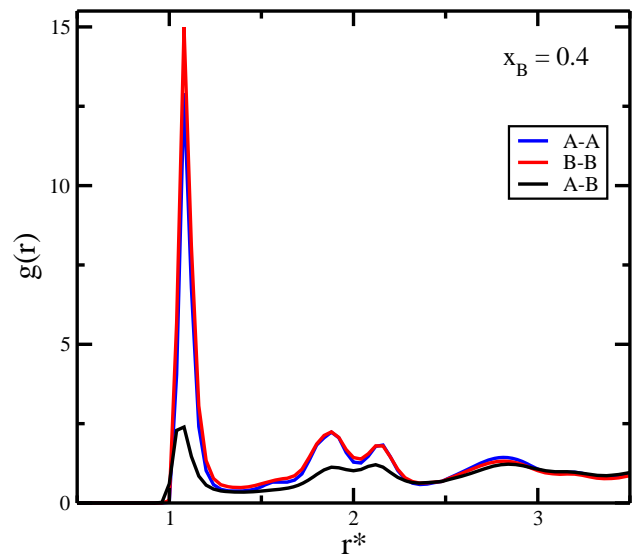

**Figure 4(b)**

**Figure 4.** Plot of partial radial distribution function of **(a)** the structure breaker parent liquid and **(b)** the corresponding inherent structure at the composition $x_B = 0.4$; parent liquid at reduced temperature, $T^* = 1.6$. Note that the peak of $g^{AB}$ (r) shows minimum in comparison with $g^{AA}$ (r) and $g^{BB}$ (r).

# VII. Correlation between instantaneous molecular arrangement in the parent liquid and the corresponding inherent structure

Although the potential energy landscape itself is independent of temperature, the manner of its exploration by the system is temperature dependent. This is reflected in the inherent structures populated by the system at different temperatures. We investigate the temperature dependence of instantaneous molecular arrangements by obtaining the inherent structures for a parent binary liquid at low temp. $T^* = 1.6$ as well as at an elevated temp. $T^* = 5.0$ with mole fraction $x_B = 0.4$. As discussed earlier, some of these structures are manifestations of the energy-entropy compensation rule. The higher entropy and higher energy states are those where the particles are more dispersed while the lower entropy and lower energy states are those which contain seeds of phase separation by dint of already containing molecular packing where A particles are surrounded by more A than B particles. We show below examples of various structural patterns at two different temperatures.

### (A) Snapshots at mole fraction $x_B = 0.4$ and temperature $T^* = 1.6$

In order to understand the nature of phase separation as well as to capture the signatures of spinodal decomposition, we have taken several snapshots of parent liquid and corresponding inherent structure at reduced temperature $T^* = 1.6$, as shown in **Figure 5** below. **Figure 5 (a)** shows an instantaneous parent structure while **Figures 5 (b), (c), (d), (e)** all four depict molecular arrangements for inherent structure shown from different angles. While **Figure 5 (a)** shows no large scale phase separation, **Figures 5 (b), (c), (d), (e)** show phase separation between constituent atoms.

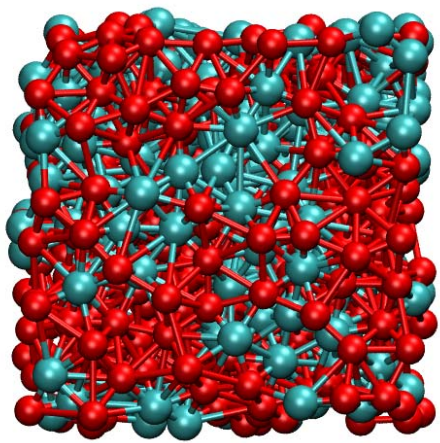
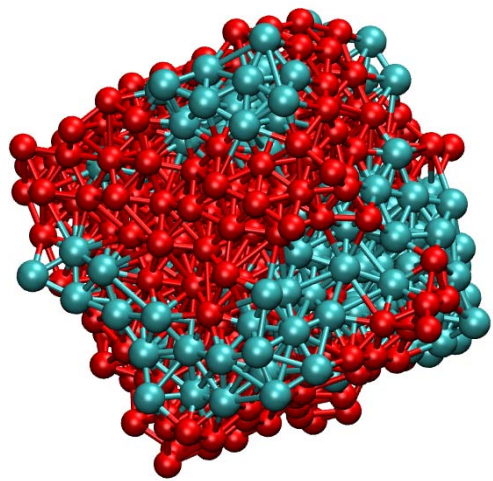

**Figure 5(a)**                    **Figure 5(b)**

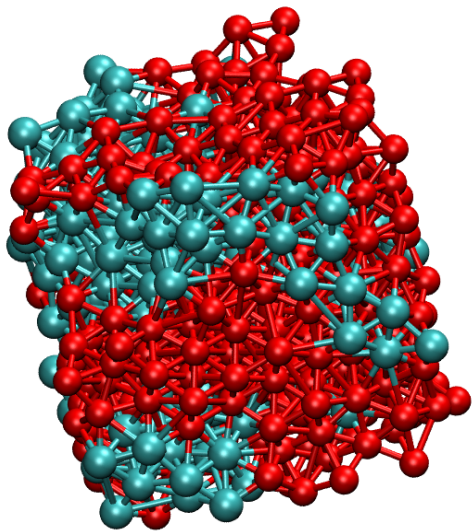
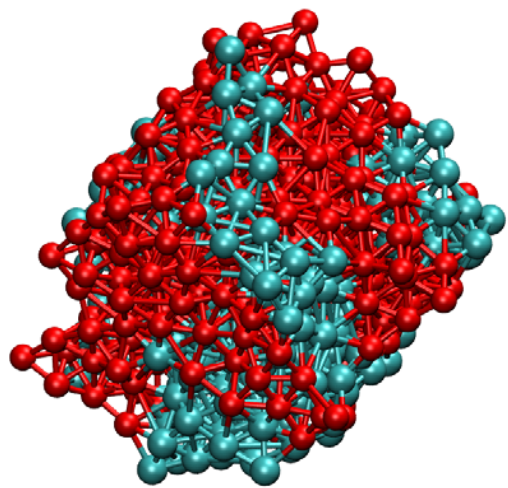

**Figure 5(c)**                    **Figure 5(d)**

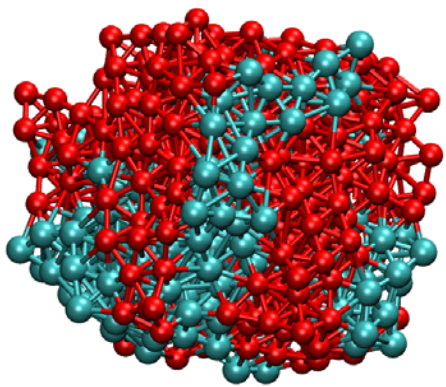

**Figure 5(e)**

**Figure 5(a).** Snapshot for **parent liquid** with n = 500 at solute composition $x_B$ = 0.4 and temperature T* = 1.6

**(b), (c), (d), (e).** Snapshots for the corresponding inherent structure in CPK mode of drawing method in graphical representation where red colour signifies solvent (A) and blue for solute (B). Note that the snapshots are taken from different angles.

**(B)  Snapshots at mole fraction $x_B$ = 0.4 and temperature T* = 5.0**

We show several snapshots of parent and corresponding inherent structure at comparatively higher temperature, T* = 5.0, as shown in **Figure 6** below. **Figure 6(a)** shows an instantaneous parent structure which seems to be almost homogeneous. **Figures 6(b), (c), (d), (e)** all four depict molecular arrangements of corresponding inherent structure taken from different angles and these four snapshots show phase separation between constituent atoms.

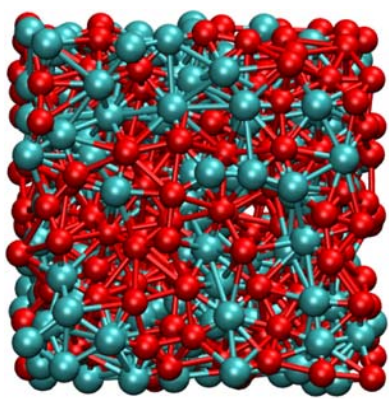

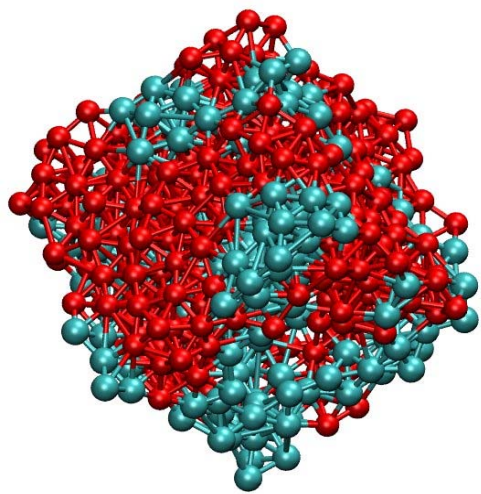

**Figure 6(a)**          **Figure 6(b)**

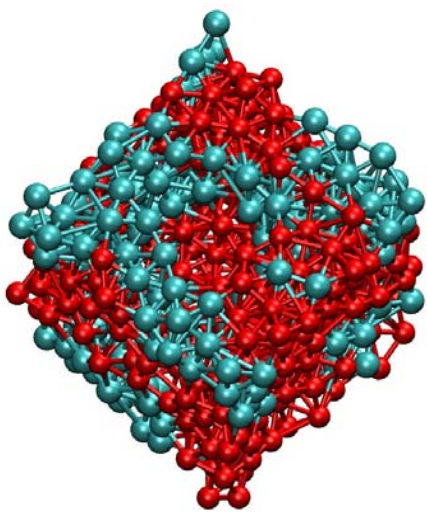

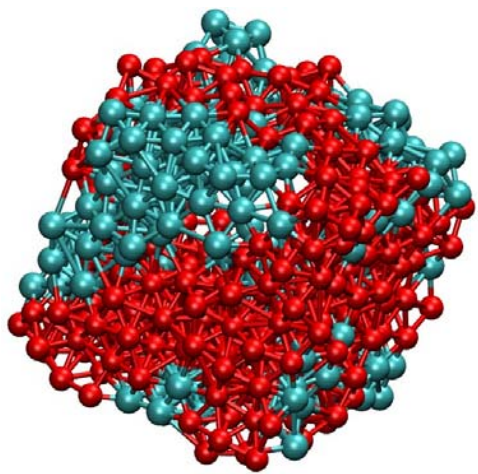

**Figure 6(c)**          **Figure 6(d)**

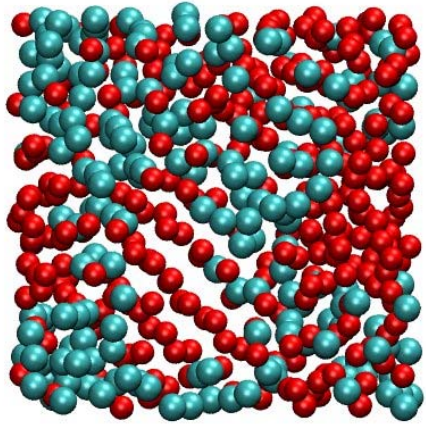

**Figure 6(e)**

**Figure 6.** Snapshots at solute composition $x_B = 0.4$ **(a)** Parent liquid with n = 500 and temperature T* = 5.0 **(b), (c), (d).** Corresponding inherent structure in CPK mode where red colour signifies solvent (A) and blue for solute (B). Note that the snapshots are taken from different angles. **(e)** Same snapshot but taken in VDW mode. Note the beautiful string-like structure as in **Figure 6(e)**.

Presence of inter-connected regions in the inherent structure suggests that such potential minima in the configuration arrangement of molecules in the parent phase (which is homogeneous on a long time and length scales) could provide a driving force for the phase separation through spinodal decomposition at low temperatures. In the linear theory, onset of spinodal decomposition is signaled by the presence of a critical wavelength $\lambda^*$. Composition fluctuations of wavelengths larger than this critical wavelength are predicted to become unstable and lead to phase separation by spinodal decomposition. In the linear theory, the value of this wavelength is governed by the equation,

$$\lambda^* = 2\pi\sqrt{\frac{K}{f''}}L, \tag{4}$$

where λ* is the wavelength, $K$ is the measure of surface tension or surface rigidity, $f$ is the composition dependent free energy density of the homogeneous system, and $f''$ is the second derivative at the free energy at maximum when the spinodal decomposition occurs [15-19]. This equation provides a quantitative dependence of the width of the spinodal strip on surface tension and the barrier curvature.

However, it appears that the linear theory is incapable of capturing the diversity of patterns that we observe in the simulations. It also appears that *the microscopic details of the patterns* observed at the intermediate times depend on the instantaneous composition of the homogeneous phase. This effect has not yet been included in the existing theories. Note that the coarsening and the *average phase separation behavior* need not depend on the initial details of the configuration. To elaborate on this issue, the linear theory employs a van der Waals (or, Landau) type free energy functional of the position dependent concentration. This average description cannot include the effects of microscopic details that control phase separation at intermediate times. Description of such patterns requires a more elaborate theoretical description in terms of more microscopic order parameter than just the composition fluctuation. At present such a description is lacking. It might be highly non-trivial to develop a theory to capture the observed dependence of the pattern on the inherent structure energy distribution. One would certainly need to include an enhanced set of order parameters, like we needed to describe the nucleation at large metastability [32].

To illustrate the point made above, let us note that each parent structure, even when belonging to the same free energy profile, can have quite different enthalpy and entropy. This is evident from the potential energy distribution of inherent structures. Thus, the initial states might evolve differently (in the microscopic or mesoscopic details) depending on the entropy-enthalpy balance of the initial state. That is, there is diversity in the pattern formation, while the coarsening scenario remains the same for all these particles. We refer to the elegant work by Rabani et al who addressed some of these issues [27]. In order to include such effects, it seems that we need to include the dependence of the system's dynamics on the entropy as in Adam-Gibbs expression [33] given by,

$$D = A \exp^{\frac{-B}{TS_C}},$$ (5)

where $S_C$ is the configurational entropy, A and B are constant coefficients. It will be worthwhile to develop such a theory. Expression for long wavelength fluctuations (4) and Adam-Gibbs expression (5) along with inherent structures, illustrate the competitive nature of phase separation in the binary mixtures, and should be relevant in the pattern selection.

## VIII. Conclusions

Several comments on this work are now in order.

(i) We find the average inherent structure energy shows positive deviations and viscosity shows negative deviations for structure breaker liquids. On the other hand, the average inherent structure energy (viscosity) of structure maker liquids shows negative (positive) deviations from Raoult's law (reported in supplimentary material section **S1**).

(ii) We explore that the distribution of inherent structure energy, $P ( E_{IS} )$ in the case of structure breaking liquids is *always broader* than that of structure maker liquids studied (reported in the supplementary material section **S2**) demonstrating the relatively larger role of entropy in stabilizing the structure breaking parent mixtures.

(iii) We find an interesting correlation between the energy of the inherent structure of the binary mixture and the microscopic mode of phase separation in the inherent structure. In particular, the existence of spinodal decomposition like structures in the energy landscape is quite fascinating and *has not been reported before*.

(iv) Most importantly, the inherent structure of structure breaking liquid is always phase separated, in contrast to that of a structure maker liquid where inherent structure remains homogeneous (reported in the supplementary material section **S3**). This aspect of phase separation in the inherent structure is similar to the case of liquid crystal where we found that the inherent structure of an isotropic parent phase is always nematic [11].

(v) In order to study larger system with 2048 particles we needed to employ nearest neighbor list to calculate the pattern and inherent structures. Nearest neighbor list errs in understanding the surface tension which depends critically on the range of interaction [28]. Thus, the patterns, also exhibit the same qualitative features, are a bit less sharp (reported in the supplementary material section **S4**). Hence, comparisons of results between two different system sizes (500 and 2048 particles) show no significant differences in phase separation pattern. In particular, the main conclusions regarding the spinodal decomposition in the inherent structure remain intact, except that the spinodal stripes become thicker in the larger system.

## *ACKNOWLEDGEMENT*

We thank Prof. S. Sastry for insightful discussions. We also thank B Jana, S Banerjee, R Biswas, S Roy, Dr. D. Chakrabarti, Prof. K. Chattopadhyay for helpful discussions. This work was supported in parts by grants from DST and CSIR (India). BB thanks DST for a JC Bose Fellowship.

**References:**

1. G. Srinivas, A. Mukherjee and B. Bagchi, J. Chem. Phys. **114**, 6220 (2001).

2. A. Mukherjee, G. Srinivas, and B. Bagchi, Phys. Rev. Lett. **86**, 5926 (2001).

3. S. Sastry, Phys. Rev. Lett. **85**, 590 (2000).

4. J. R. Fernandez, P. Harrowell, Phys. Rev. E **67**, 011403 (2003).

5. W. Kob and H.C. Andersen, Phys. Rev. E **51**, 4626 (1995).

6. W. Kob and H.C. Andersen, Phys. Rev. Lett. **73**, 1376 (1994).

7. F. Calvo, T. V. Bogdan, V. K. de Souza and D. J. Wales, J. Chem. Phys. **127**, 044508 (2007); T. F. Middleton, J. H. Rojas, P. N. Mortenson and D. J. Wales, Phys. Rev. B **64**, 184201 (2001).

8. A. Heuer, Phys. Rev. Lett. **78**, 4051 (1997).

9. F. H. Stillinger and T.A. Weber, Science **225**, 983 (1984); F. H. Stillinger, Science **267**, 1935 (1995); T. A. Weber and F. H. Stillinger, Phys. Rev. B **31**, 1954 (1985).

10. S. Sastry, Nature (London) **409**, 164 (2001); S. Sastry, P. G. Debenedetti, and F. H. Stillinger, Phys. Rev. E **56**, 5533 (1997); S. Sastry, P. G. Debenedetti, and F. H. Stillinger, Nature (London) **393**, 554 (1998).

11. D. Chakrabarti and B. Bagchi, PNAS **103**, 7217 (2006); D. Chakrabarti and B. Bagchi, Adv. Chem. Phys. **141**, 249 (2009).

12. D.J.Wales, *Energy Landscapes* (Cambridge University Press, Cambridge, 2003) ISBN 0521814154.

13. J. D. Bryngelson and P. G. Wolynes, J. Phys. Chem. **93**, 6902 (1989); J. N. Onuchik, P. G. Wolynes, Z. Luthey-Schulten, N. D. Socci, PNAS USA **92**, 3626 (1995); J. Bryngelson, J. Onuchik, N. Socci and P. G. Wolynes, Proteins: Struct., Funct., Genet. **21**, 167 (1995).

14. R. Zwanzig, PNAS **92**, 9801 (1995); R. Zwanzig, A. Szabo, B. Bagchi, PNAS **89**, 20 (1992); R. Zwanzig, PNAS **85**, 2029 (1988).

15. J. D. Gunton, M. San Miguel and P. S. Sahni, *Phase transitions and critical phenomena*, edited by C. Domb and J. L. Lebowitz (Academic, London), **8** (1983).

16. J. W. Cahn and J. E. Hilliard, J. Chem. Phys. **28**, 258 (1958).

17. J. S. Langer, M. Bar-on and H. D. Miller, Phys. Rev. A **11**, 1417 (1975).

18. K. Binder and D. Stauffer, Phys. Rev. Lett. **33**, 1006 (1974).

19. E. D. Siggia, Phys. Rev. A **20**, 595 (1979).

20. T. Koga and K. Kawasaki, Physica A **196,** 389 (1993).

21. Z. Mao, C. K. Sudbrack, K. E. Yoon, G. Martin and D. N. Seidman, Nature Mater. **6**, 210 (2007).

22. V. Talanquer and D. W. Oxtoby, J. Chem.Phys. **109,** 223 (1998).

23. J. F. Peters and E. S. Berney IV, J. Geotech and Geoenvir. Engrg. **136**, 310 (2010).

24. J. Wang, I L McLaughlin and M. Silbert, J. Phys.: Condens. Matter **3**, 5603 (1991).

25. P. R. ten Wolde and D. Frenkel, Science **277**, 1975 (1997).

26. M. K. Mitra and M. Muthukumar, J. Chem.Phys. **132**, 184908 (2010).

27. E. Rabani, D. R. Reichman, P. L. Geissler and L. E. Brus, Nature **426**, 271 (2003).

28. M. Santra and B. Bagchi, J. Chem. Phys. **131**, 084705 (2009).

29. W. H. Press, B. P. Flannery, S. A. Teukolsky, and W. T. Vetterling, Numerical Recipes in FORTRAN (Cambgidge University Press, Cambridge, 1990).

30. J. D. Stevenson, J. Schmalian and P. G. Wolynes, Nat. Phys. **21**, 268 (2006).

31. H. Saito, M. Yoshinaga, T. Mihara, T. Nishi, H. Jinnai, J. Phys.: Conf. Ser. **184**, 012029 (2009).

32. P. Bhimalapuram, S. Chakrabarty and B. Bagchi, Phys. Rev. Lett. **98**, 206104 (2007).

33. G. Adam and J. H. Gibbs, J. Chem. Phys. 43, **139** (1965).

# Supplementary material

## S1: Nonlinear composition dependence of inherent structure energy and its correlation with viscosity for *structure promoting* binary mixture

**Figure S1.1** depicts the variations of the average inherent structure energy as well as the viscosity with different solute compositions for structure promoting (SP) binary mixture model. The average inherent structure energy ($<E_{IS}>$) shows a minimum at solute composition $x_B = 0.4$ for structure promoter. This signifies that the system is more structured in the configuration space and there is a slowing down of dynamics at this composition as the system is ordered here. The viscosity exhibits exactly the opposite composition dependence, in agreement with the physical explanation discussed above.

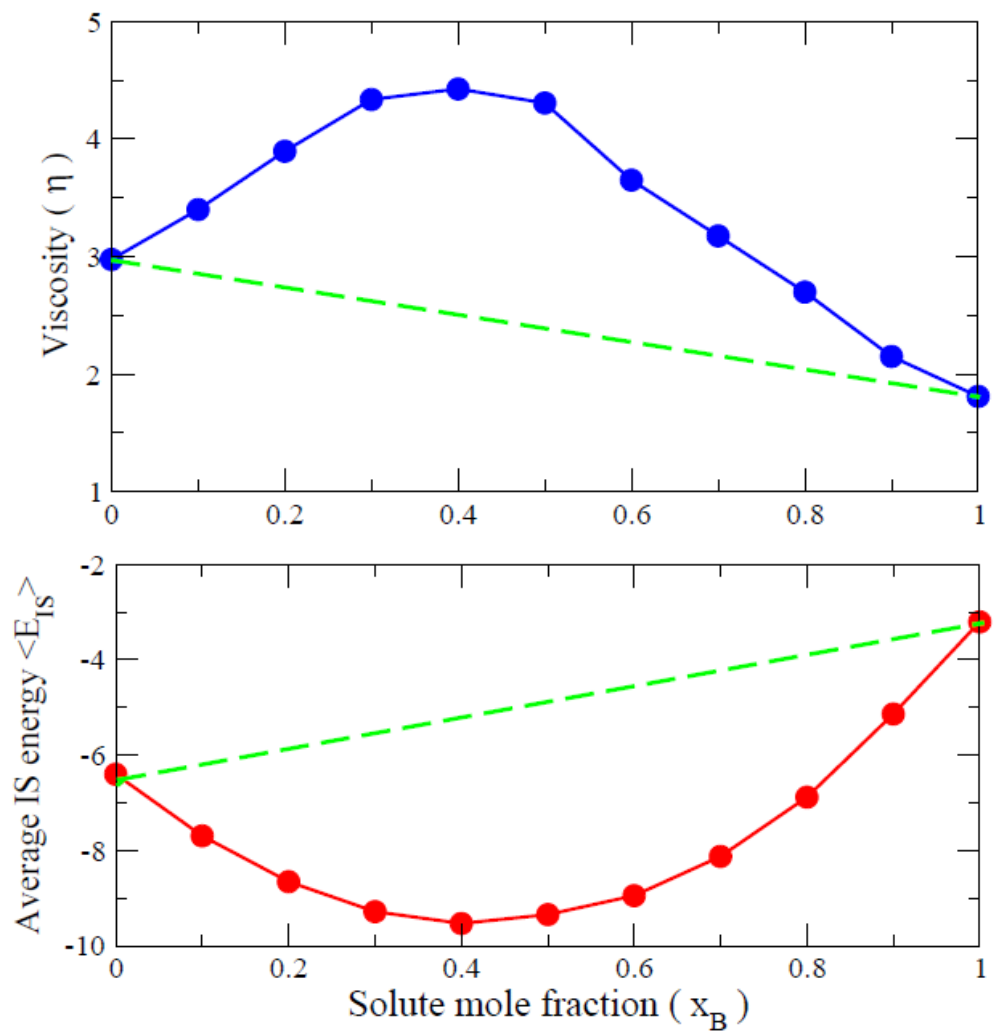

**Figure S1.1**. Plot of the computed viscosity and the average inherent structure energy at different solute mole fractions for structure promoting binary mixture.

# S2: Comparison between inherent structure energy distributions of structure *promoting* and structure *breaking* binary mixtures

**Figures S2.1(a) and S2.1(b)** show the distributions of inherent structure energies for the structure promoting (SP) and structure breaking (SB) model respectively, as obtained from our MD simulations. The distribution curve for structure breaking binary mixture is broader/wider where as the plot for structure promoter is comparatively narrower. We have attempted to fit these curves with Gaussian distribution. While distribution of the inherent structure energies of structure promoter can be fitted to a Gaussian distribution, the same is not true for structure breaker. The inherent structure energy distribution remained noisy even after averaging over larger number (2000) of configurations for structure breaker.

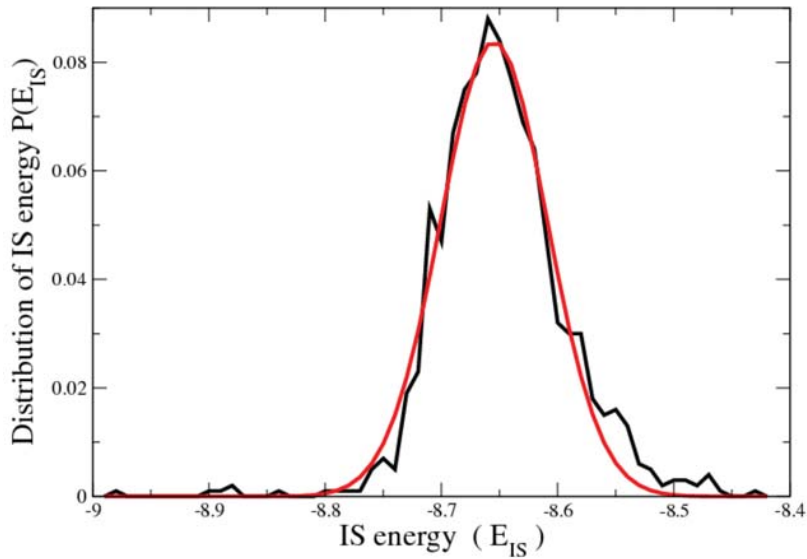

**Figure S2.1.  (a)** Distribution of inherent structure energy for *structure promoting* model corresponds to parent liquid with temp T*= 1.0 and solute mole fraction $x_B = 0.2$. The distribution curve is fitted with Gaussian distribution. Here red colour represents Gaussian distribution and black for distribution of inherent structure energy.

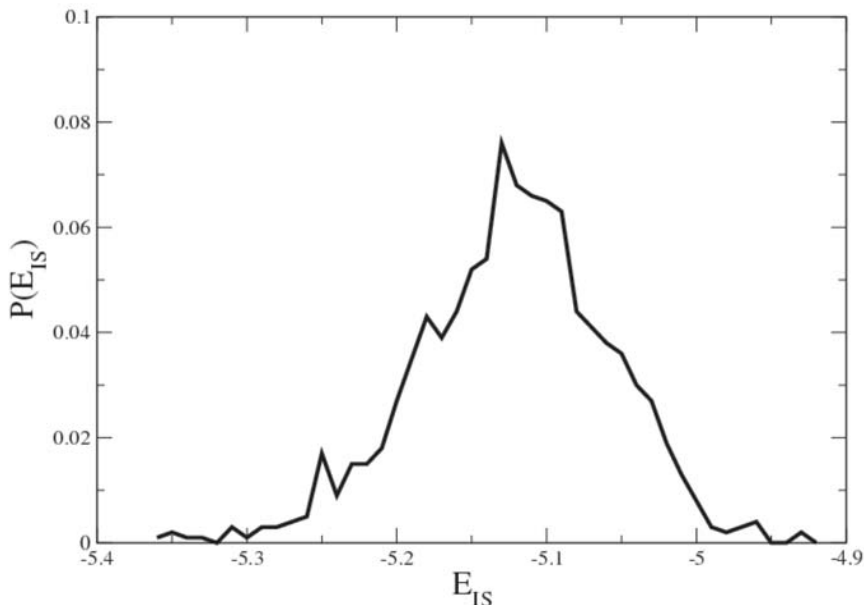

**Figure S2.1. (b)** Distribution of inherent structure energy for *structure breaking* model corresponds to parent liquid at mole fraction ( $x_B$ ) = 0.2 and temperature T* = 1.6. Please note that this distribution could not be fitted to a Gaussian form.

The broader inherent structure energy distribution for structure breaking model clearly indicates larger entropic contribution towards the stability and homogeneity of structure breaker than structure promoting liquid. On the other hand, the average energy of the inherent structure of structure promoting model is significantly smaller than that for structure breaking model, indicating a larger enthalpic stabilization for structure promoter. Therefore, we can conclude that the homogeneous state of structure breaking solute/solvent system is stabilized by entropy while that of structure promoter is stabilized by enthalpy.

S3: **Snapshots for parent structure of *structure promoting* binary mixture and its corresponding inherent structure**

**Figure S3.1(a)** represents the snapshot for parent liquid of structure promoting binary mixture with solute mole fraction $x_B = 0.4$ at temperature $T^* = 1.0$ and **S3.1(b)** represents its corresponding inherent structure. These snapshots show there are no significant changes in the pattern of molecular arrangements in parent and its corresponding inherent structure for structure promoting binary mixture.

**Figure S3 .1(a)**                    **Figure S3.1(b)**

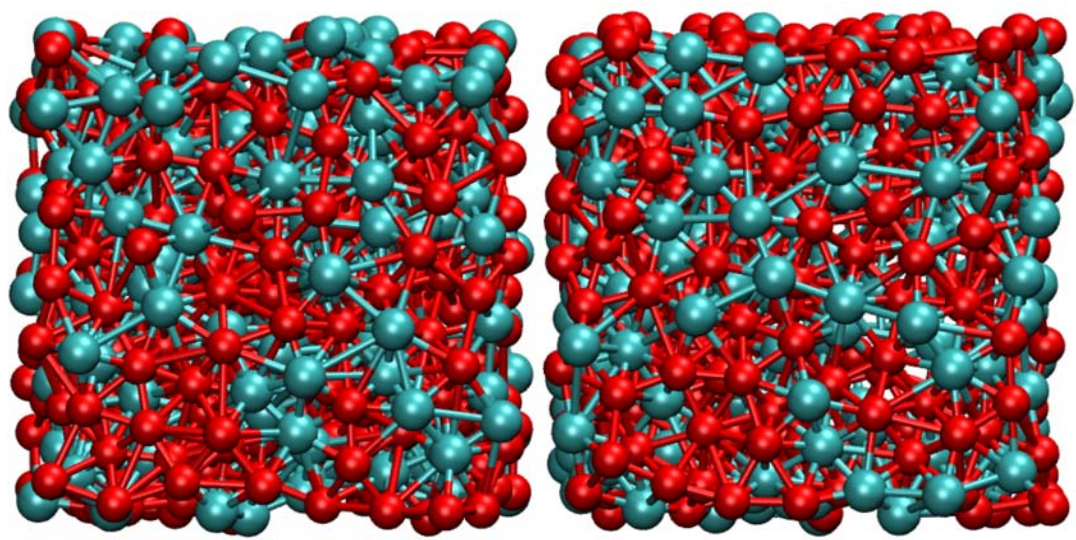

**Figure S3.1**. **(a)** Snapshot for parent structure of structure promoting binary mixture with n = 500 at solute composition $x_B = 0.4$ and at temperature $T^* = 1.0$. **(b)** Snapshot for corresponding inherent structure

# S4: Snapshots of parent liquid and corresponding inherent structure at mole fraction $x_B = 0.4$ and temperature $T^* = 1.6$ for larger system (n=2048) of structure breaking binary mixture

**Figure S4.1(a)** represents the snapshot for parent liquid of larger system (n=2048) of structure breaking binary mixture with solute mole fraction $x_B = 0.4$ at temperature $T^* = 1.6$ and **S4.1(b), (c)** represent its corresponding inherent structure. The snapshots of inherent structure for structure breaking binary mixture show signature of spinodal decomposition.

**Figure S4 .1(a)**

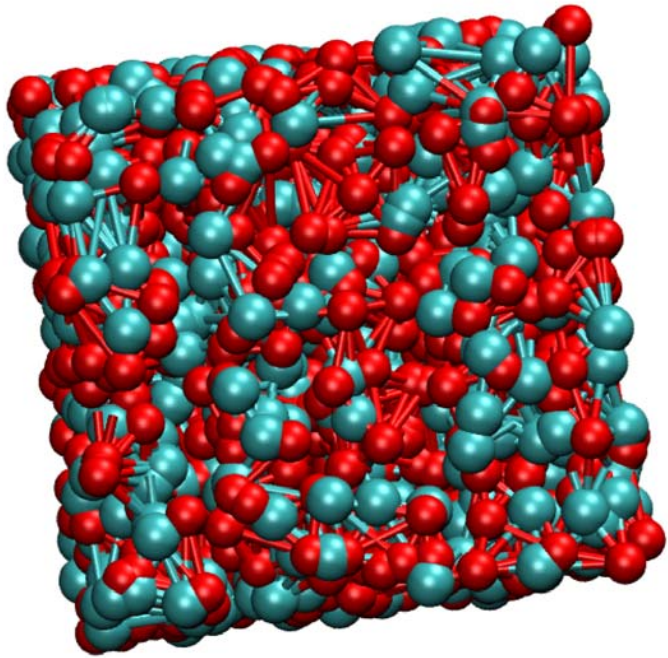

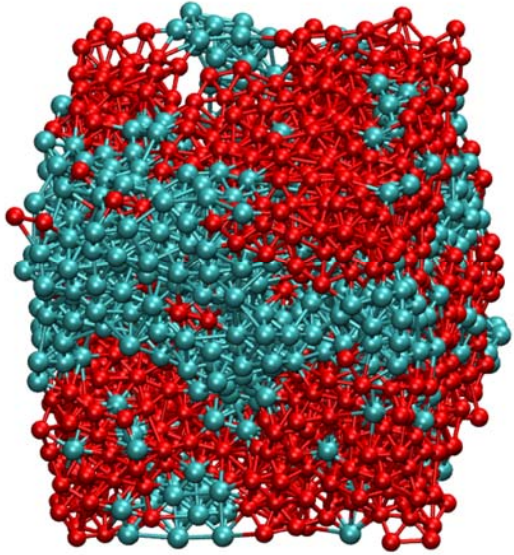
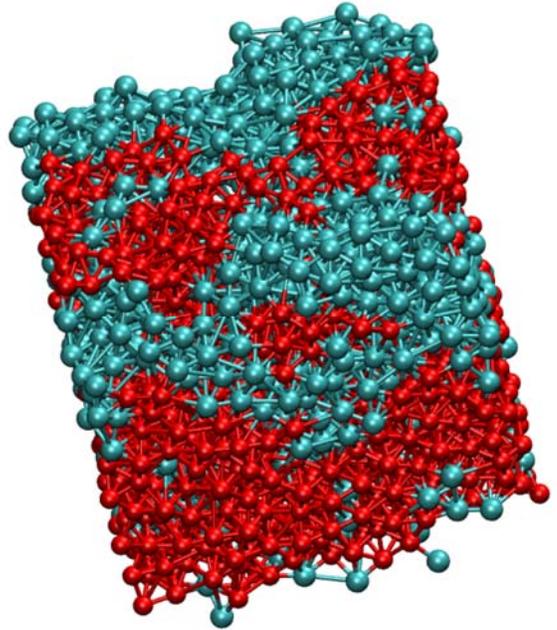

**Figure S4.1**. **(a)** Snapshot for parent structure of *structure breaking* liquid with n = 2048 at solute composition $x_B$ = 0.4 and at temperature T* = 1.6. **(b), (c)** Snapshots for corresponding inherent structure. Note the snapshots are taken from different angles.